\documentclass [prb,twocolumn,superscriptaddress,showpacs,preprintnumbers] {revtex4}
\usepackage[]{amssymb,amsmath,amsfonts}
\usepackage[]{graphicx}

\begin{document}
\title{Pseudo-surface acoustic waves in hypersonic surface phononic crystals}

\author{D. Nardi}
\affiliation{Dipartimento di Matematica e Fisica, Universit\`a Cattolica del Sacro Cuore, I-25121 Brescia, Italy}
\affiliation{Dipartimento di Fisica, Universit\`a degli Studi di Milano, I-20122 Milano, Italy}

\author{F. Banfi} \email[]{francesco.banfi@dmf.unicatt.it}
\author{C. Giannetti}
\affiliation{Dipartimento di Matematica e Fisica, Universit\`a Cattolica del Sacro Cuore, I-25121 Brescia, Italy}

\author{B. Revaz}
\affiliation{\'Ecole Polytechnique F\'ed\'erale de Lausanne, Laboratoire de microsyst\`emes 3, CH-1015 Lausanne, Switzerland}

\author{G. Ferrini}
\affiliation{Dipartimento di Matematica e Fisica, Universit\`a Cattolica del Sacro Cuore, I-25121 Brescia, Italy}

\author{F. Parmigiani}
\affiliation{Dipartimento di Fisica, Universit\`a degli Studi di Trieste and Sincrotrone Trieste, I-34012 Basovizza, Trieste, Italy}

\date{\today}

\begin{abstract}
We present a theoretical framework allowing to properly address the nature of surface-like eigenmodes in a hypersonic surface phononic crystal, a composite structure made of periodic metal stripes of nanometer size and periodicity of 1~$\mu$m, deposited over a semi-infinite silicon substrate. In surface-based phononic crystals there is no distinction between the eigenmodes of the periodically nanostructured overlayer and the surface acoustic modes of the semi-infinite substrate, the solution of the elastic equation being a pseudo-surface acoustic wave partially localized on the nanostructures and radiating energy into the bulk. This problem is particularly severe in the hypersonic frequency range, where semi-infinite substrate's surface acoustic modes strongly couple to the periodic overlayer, thus preventing any perturbative approach. We solve the problem introducing a surface-likeness coefficient as a tool allowing to find pseudo-surface acoustic waves and to calculate their line shapes. Having accessed the pseudo-surface modes of the composite structure, the same theoretical frame allows reporting on the gap opening in the now well-defined pseudo-SAW frequency spectrum. We show how the filling fraction, mass loading and geometric factors affect both the frequency gap, and how the mechanical energy is scattered out of the surface waveguiding modes.
\end{abstract}

\pacs{62.25.-g, 68.35.Iv, 43.35.+d, 43.35.Pt}
%

\maketitle

\section{Introduction}
\indent The idea of creating solids with artificial modulation in their physical parameters\cite{Esaki} has proved fruitful both in the frame of coherent electronic transport and photonics,\cite{Knight, Russell2003} with the advent of semiconductor superlattices and photonic crystals,\cite{Yablonovitch, John, Krauss} respectively. The analogy between photons and phonons suggested considering phononic crystals,\cite{Kushwaha} periodic elastic composites of two or more vibrating materials, the interest being triggered by the possibility of achieving a frequency gap in the elastic modes dispersion relations and the wealth of applications stemming from tailoring the gap itself.\\
\indent Extensive investigation of acoustic band structures of periodic elastic composite has been carried out on two-dimensional phononic crystals both infinite\cite{Kushwaha} and surface-terminated.\cite{Tanaka1998, Tanaka1999} The archetypal 2D structure consists of parallel rods embedded in an elastic background. Rods can be actual or virtual ones (drilled holes). In the real case of a surface-terminated phononic crystal, an interesting class of acoustic modes, addressed as surface acoustic waves (SAWs) - arising from breaking of the translational symmetry when passing from an infinite to a semi-infinite medium - propagate confined to the elastic medium surface, the penetration depth being of the order of their spatial period. Recent works elucidated some of the characteristics of surface acoustic waves in 2D phononic crystals of different geometrical configurations.\cite{Haus, Chen, Datta, Robinson}\\
\indent The present paper focuses on pseudo-surface acoustic modes, reminiscent of surface acoustic waves, in surface-based phononic crystals. These structures are obtained patterning the surface of a substrate material either by metal deposition or substrate etching. The influence of a periodically structured overlayer on the acoustic-field eigenmodes of a thick or thin homogeneous slab has been studied by many research groups and effects such as band folding,\cite{Dutcher1992, Maznev2009} mode leakage,\cite{Glass1983} opening of frequency gaps,\cite{Dutcher1992, Glass1983, Zhang2006, Profunser2006, Bonello2007, Maznev2009} and interaction between slab and overlayer modes\cite{Sainidou2006, Bonello2007, Liu2008} have been discussed. Contrary to 2D phononic crystals, for surface acoustic modes in surface-based phononic crystals there is no distinction between the eigenmodes of the nanostructures and SAWs, the solution of the elastic equation being a pseudo-surface acoustic wave partially localized on the nanostructures and radiating energy into the bulk. This issue is particularly severe in the hypersonic frequency range, where the semi-infinite substrate's surface acoustic modes strongly couple to the periodic overlayer, thus preventing any perturbative approach. In this physical scenario, we present a theoretical framework allowing to properly address the nature of surface-like eigenmodes in a hypersonic surface phononic crystal. We solve the problem introducing a surface-likeness coefficient as a tool allowing to find pseudo-surface acoustic waves and to calculate their line shapes. Having accessed the pseudo-surface modes of the composite structure, the same theoretical frame allows reporting on the gap opening in the now well-defined pseudo-SAW frequency spectrum. We show how the filling fraction, mass loading and geometric factors affect both the frequency gap, and how the mechanical energy is scattered out of the surface waveguiding modes.\\
\indent Our results are of interest also from an applicative stand-point. The quest for hypersonic acoustic waveguides,\cite{Morvan} sources of ultrafast coherent acoustic waves and nano-opto-acoustic transducers in general operating at ever higher frequencies, requires patterning periodic structures of ever decreasing periodicities. In this frame, surface-based phononic crystals provide a technological advantage over 2D phononic ones, the processing technique involving the surface only, being easily scalable below 100~nm and suitable for high frequency transducers technology. The theoretical tool here presented, together with the comprehension of how the construction parameters affect the frequency gap and surface waveguiding mode, will be a valuable tool for inspecting the pseudo-acoustic modes relevant for applications in view of device engineering beyond a trial-and-error approach.\\
\indent The present work is organized as follows: in Sec.~\ref{Mechanical_model} we outline and solve the full mechanical problem, calculating the vibrational normal modes of the composite system. The presence of periodic nickel stripes modifies the properties of surface waves.\cite{Landau, Auld, Cheng} The SAW is no longer an eigenmode of the elastic eigenvalue problem. Proper assessment of a pseudo-SAW in a surface-based phononic crystal is achieved in Sec.~\ref{Pseudo-SAW}. We introduce a SAW-likeness coefficient permitting to discriminate pseudo-SAWs among the entire set of eigenmodes and to calculate their line-profiles. Investigation of the gap opening in a surface-based phononic crystal as a function of the filling fraction $p$, beyond a perturbative approach, is presented in Sec.~\ref{Frequency_gap}. We show how the frequency gap depends on the set of geometric factors $\left\{p,h\right\}$ and mass loading $\left\{m,\rho\right\}$ - $m$, $\rho$ and $h$ being the nickel stripe's mass, density and height, respectively. In Sec.~\ref{Mechanical_energy} we analyze the energy distribution in the system as a function of $p$ and relate it to the scattering of SAWs into the bulk.
\section{Mechanical model} \label{Mechanical_model}
\begin{figure}[t]
\centering
\includegraphics[bb= 138 181 444 268,keepaspectratio,clip,width=0.9\columnwidth]{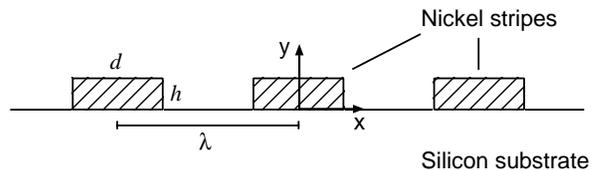}
\caption{Schematic diagram of the surface phononic crystal. Isotropic nickel stripes of width $d$ and height $h$ are deposited on a crystalline silicon substrate. The grating has period $\lambda$. Propagation direction is along the $x$ axis, surface normal to the bulk is along the $y$ axis.} \label{Phononic_crystal}
\end{figure}

\indent We assume the system to be an elastic continuum composed of a periodic metallic grating deposited on a silicon substrate. In Fig.~\ref{Phononic_crystal} we consider the general configuration where elastic isotropic nickel stripes of width $d$ and height $h=50$ nm are deposited on a crystalline silicon substrate. The grating has period $\lambda=1$~$\mu$m, thus ensuring pseudo-SAWs in the hypersonic range. Calculations are performed increasing $d$, in order to cover the entire filling fraction $p$ range $(0,1)$ - where $p=d/\lambda$ - and explore different regimes, from a perturbative one ($p\ll1$) to substrate full coverage ($p=1$). The acoustic equation of motion governing the displacement $\textbf{u}(\textbf{r},t)$ of the composite system is
\begin{equation}
\partial_j\left[c_{ijmn}\left(\textbf{r}\right)\partial_n u_m\right]=\rho\left(\textbf{r}\right)\ddot{u}_i\;,
\label{Acoustic_equation1}
\end{equation}
where $\rho\left(\textbf{r}\right)$ and $c_{ijmn}\left(\textbf{r}\right)$ are the position dependent mass density and elastic stiffness tensor, respectively, and the summation convention over repeated indices is assumed. For an harmonic time dependence $e^{i\omega t}+c.c.$, Eq.~\ref{Acoustic_equation1} reads
\begin{equation}
\partial_j\left[c_{ijmn}\left(\textbf{r}\right)\partial_n u_m\right]=-\rho\left(\textbf{r}\right)\omega^2 u_i\;.
\label{Acoustic_equation2}
\end{equation}
\indent We solve the eigenvalue problem via the finite elements method.\cite{Comsol} As shown in Fig.~\ref{SAW_solutions}(a), the model consists of a two dimensional silicon rectangular cell with a nickel stripe on top.
\begin{table}[b]
\begin{center}
\centering \caption{\label{Material_properties} Material properties for Si substrate\cite{Okhotin} and Ni stripes:\cite{Davies} Young's modulus~$E$, Poisson's ratio~$\sigma$ and mass density~$\rho$.\\}
\begin{ruledtabular}
\begin{tabular}{lccc}
 & $E$ (GPa) & $\sigma$ & $\rho$ (Kg/m$^{3}$)\\
\hline
Si substrate & 131 & 0.27 & 2330\\
Ni stripes & 219 & 0.31 & 8900\\
\end{tabular}
\end{ruledtabular}
\end{center}
\end{table}
The material properties for the silicon substrate and the nickel stripes, that enter in the expression of the elastic stiffness tensor, are reported in Table~\ref{Material_properties}. The silicon substrate's crystalline orientation is accounted for in the expression of the elastic stiffness tensor and, in the present case, the $x$ axis is taken along the Si(100) crystalline direction. To reproduce the entire nanostructured composite from the single unit cell, the displacements $\textbf{u}_{1}$ and $\textbf{u}_{2}$, calculated respectively on side 1 and 2 of the cell (see Fig.~\ref{SAW_solutions}(a)), are related by $\textbf{u}_{1}=e^{iK_{x,n}\lambda}\:\textbf{u}_{2}$, as required by the Bloch theorem, where $K_{x,n}=k_{x}+2\pi n/\lambda$ and $k_{x}\in\left(-\pi/\lambda,\;\pi/\lambda\right)$. The displacement is fixed to zero on the base boundary and the height $L$ of the cell is set to 100 $\mu$m, two orders of magnitude greater than the system's periodicity. This condition is required to achieve a density of states fine enough to appreciate the surface-confined system's normal modes also for $p\ll1$ and $p\sim1$, the eigenfrequencies spacing scaling as $1/L$. Attention is devoted in choosing the finesse of the quadratic ordered mesh, enabling accurate displacements calculations despite: (a) the size discrepancy between the nanostructure and the silicon cell; (b) the two orders of magnitude difference between $L$ and $\lambda$. Unless otherwise stated, the calculations are performed with $n=1$ and $k_{x}=0$ ($K_{x,1}=2\pi/\lambda$), corresponding to the first harmonic at the center of the surface Brillouin zone, calculation of the eigenfrequencies as a function of the wave vector parallel to the surface of the substrate not being the scope of the paper.
\section{Pseudo-SAW} \label{Pseudo-SAW}
\begin{figure}[t]
\centering
\includegraphics[bb= 0 0 330 570,keepaspectratio,clip,width=0.89\columnwidth]{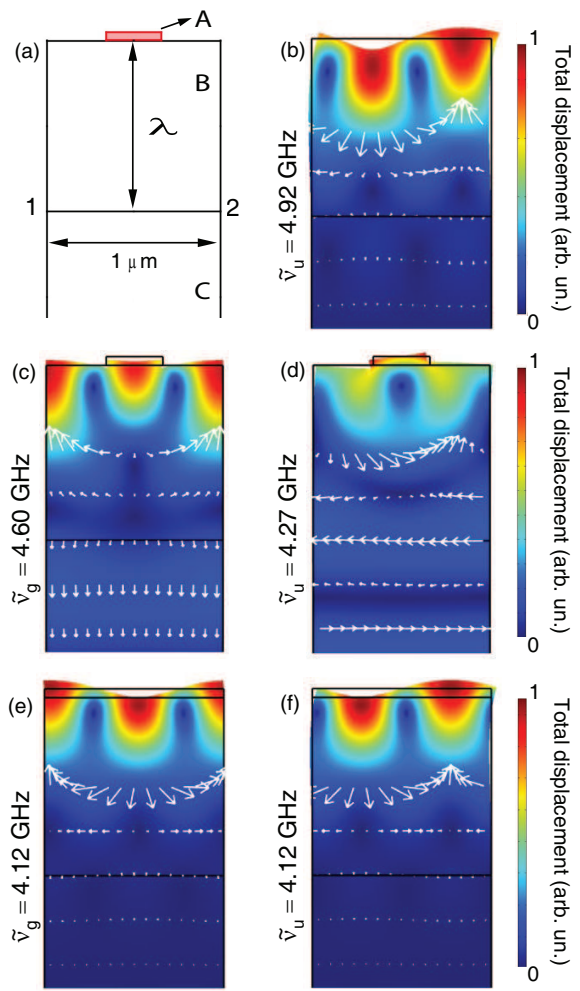}
\caption{(Color online) Nickel stripe on silicon substrate. (a)~The top 2~$\mu$m portion of the 2D rectangular unit cell is reported. The cell geometry is divided in three parts: A. Ni stripe, B. 1~$\mu$m top portion of Si substrate, C. 99~$\mu$m thick Si bulk portion. Bloch boundary conditions are set on sides 1 and 2 of the substrate. The periodicity is $\lambda=1$~$\mu$m. The height of the Si cell is 100~$\mu$m. The Ni stripe is $h=50$~nm high and, in this figure, $d=320$~nm wide. In Figures~(b) trough~(f), the deformation and arrows correspond to the displacement field, while the color scale refers to the normalized total displacement. (b)~Sin-like SAW solution for a pure silicon slab. The calculated two-fold degenerate SAW eigenfrequency is $\tilde{\nu}=4.92$~GHz. The orthogonal cos-like solution is not reported. (c)~Cos-like and (d)~sin-like pseudo-SAW solutions for $d=320$~nm. The two pseudo-SAW eigenfrequencies are $\tilde{\nu}_{g}=4.60$~GHz and $\tilde{\nu}_{u}=4.27$~GHz. (e)~Cos-like and (f)~sin-like SAW solutions for substrate full coverage. The two-fold degenerate calculated SAW eigenfrequency is $\tilde{\nu}=4.12$~GHz.}
\label{SAW_solutions}
\end{figure}

\begin{figure}[t]
\centering
\includegraphics[bb= 122 89 512 770,keepaspectratio,clip,width=0.95\columnwidth]{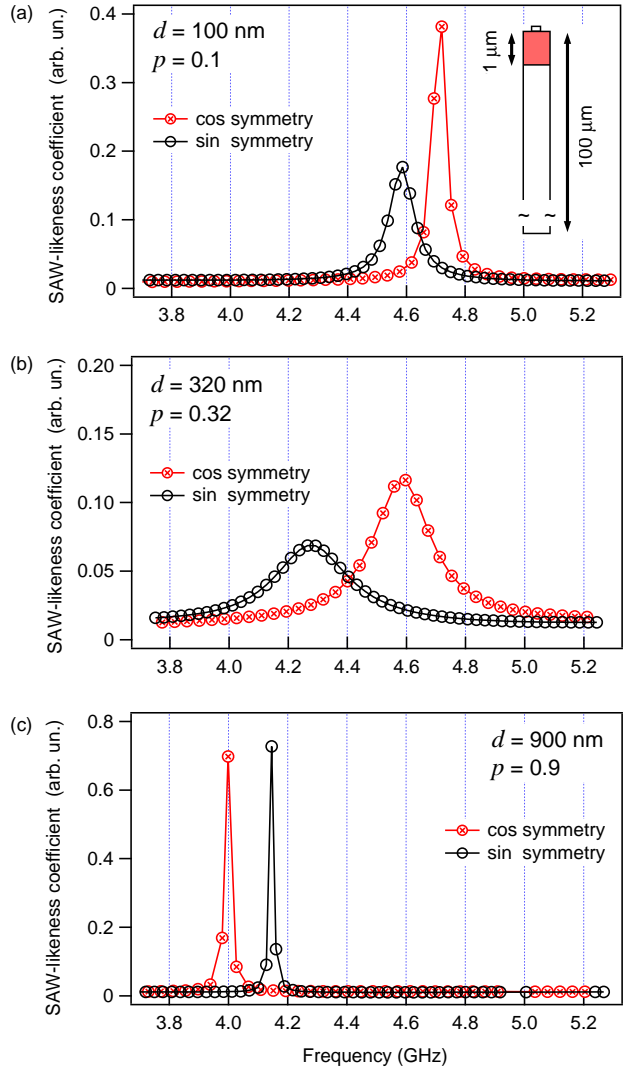}
\caption{(Color online) SAW-likeness coefficient $\alpha\left(\nu\right)$ versus the calculated eigenfrequencies. Two distributions arise, $\alpha_u\left(\nu\right)$ and $\alpha_g\left(\nu\right)$, corresponding to sin-like (empty circles) and cos-like (crossed circles) displacement profiles. Calculation for configurations with different filling fraction are reported: (a) $p=0.1$, (b) $p=0.32$ and (c) $p=0.9$. Solid lines are intended as a guide to the eye.}
\label{SAW_likeness}
\end{figure}

\begin{figure}[t]
\centering
\includegraphics[bb= 102 87 515 407,keepaspectratio,clip,width=0.9\columnwidth]{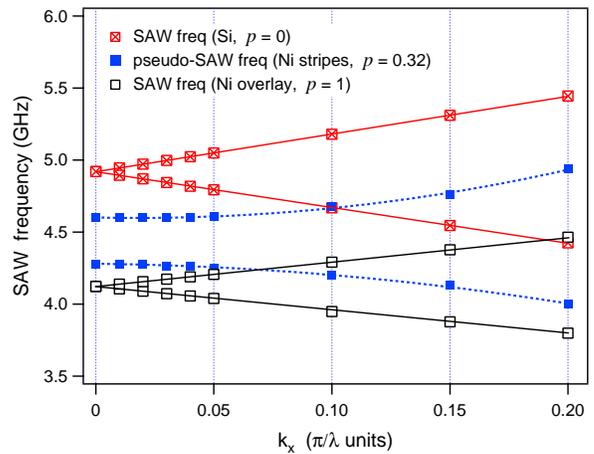}
\caption{(Color online) Dispersion relations of the elastic surface modes for small values of $k_x$ (in $\pi/\lambda$ units) and $n=1$, for a pure silicon slab ($p=0$), for silicon full coverage ($p=1$) and for the configuration with Ni stripes over Si substrate ($p=0.32$). Solid lines are fits to data. Dotted lines are a guide to the eye.}
\label{SAW_dispersion}
\end{figure}

\indent On the contrary to bulk modes, SAWs are confined in a thin layer starting at the surface of the substrate. As seen in Fig.~\ref{SAW_solutions}(b), in the case of a pure silicon slab, without stripes on top, the displacement field vanishes within a depth of the order of 1~$\mu$m, i.e. of the SAW wavelength, corresponding to the chosen grating period $\lambda$. The total displacement color scale normalized over its maximum value is reported, together with the displacement field. These modes are the elastic analogues of the electronic surface states\cite{Hufner} or the electromagnetic evanescent waves encountered at the surface of a semi-infinite solid.\cite{Courjon} The surface wave is an exact solution of the eigenvalue problem, two-fold degenerate under translational symmetry. The two solutions $\textbf{u}_u$ and $\textbf{u}_g$ have, respectively, sin (\textit{ungerade}) and cos (\textit{gerade}) symmetry. In Fig.~\ref{SAW_solutions}(b) only the sin symmetry solution is reported. The calculation of the two-fold degenerate eigenfrequency at the Brillouin zone center ($n=1$) gives $\tilde{\nu}=4.92$~GHz, in agreement with data for a SAW propagating along Si(100).\cite{Auld}\\
\indent On a uniform flat surface a surface wave does not radiate into the bulk and has a penetration depth $\sim\lambda$.\cite{Landau} The presence of periodic nanostructures is expected to modify the properties of the SAW, perturbing the stress and velocity fields associated with SAW propagation.\cite{Lin, Antonelli, Giannetti} In the periodic composite system the SAW modes are not solutions of the eigenvalue problem. The eigenmodes of the full structure, whose displacement field most closely resemble the unperturbed SAWs, are shown in Fig.~\ref{SAW_solutions}(c) and Fig.~\ref{SAW_solutions}(d). The two-fold degeneracy is lifted and a band gap opens: one mode, with sin symmetry, is found at $\tilde{\nu}_u=4.27$~GHz, whereas a second mode, with cos symmetry, is found at $\tilde{\nu}_g=4.60$~GHz. The calculated displacement field is Rayleigh-like in region B - evident from the alike displacement fields of sin-like solutions in region B of Fig.~\ref{SAW_solutions}(b), Fig.~\ref{SAW_solutions}(d) and Fig.~\ref{SAW_solutions}(f) - and bulk-like in region C - clear from the field distribution in region C of Fig.~\ref{SAW_solutions}(c) and Fig.~\ref{SAW_solutions}(d). We address these modes as pseudo-SAWs. Periodic nanostructures force the previously unperturbed surface waves to radiate elastic energy into the bulk. The analogue in the electromagnetic case, is diffraction of surface plasmons into far field by a periodic metallic grating deposited on the substrate surface. In the acoustic case, the stress at the Ni-Si interface, needed to force the stripes to follow the motion of the surface, acts as the source of energy radiation into the bulk, as explained by Lin et al.\cite{Lin} In terms of scattering of unperturbed SAWs, this effect can be rationalized as coupling of the free surface modes to the silicon bulk modes.\\
\indent The eigenmodes for the case of full substrate coverage ($p=1$) are reported in Fig.~\ref{SAW_solutions}(e) and Fig.~\ref{SAW_solutions}(f). The band gap closes, the degeneracy is recovered with $\tilde{\nu}=4.12$~GHz, and the eigenvectors are SAWs, hence no energy is radiated in the bulk. The solution is similar to that reported in Fig.~\ref{SAW_solutions}(b) for $p=0$; the overlay down-shifts the value of $\tilde{\nu}$.\\
\indent In the previous discussion, the concept of pseudo-SAW was suggested on the basis of similarities between its displacement field and that of a pure SAW. Nevertheless the theory lacks a formal definition for a pseudo-SAW, or a procedure allowing to spot out such solutions from the infinite eigenvalues set satisfying Eq.~\ref{Acoustic_equation2} in the present geometry. With this aim in mind, we define the SAW-likeness coefficient as
\begin{equation}
\alpha(\nu)\doteq\frac{\left\langle E_{A}(\nu)\right\rangle+\left\langle E_{B}(\nu)\right\rangle}{\left\langle E_{tot}(\nu)\right\rangle}\;,
\label{SAW-likeness_coefficient}
\end{equation}
where $\left\langle E_{A}(\nu)\right\rangle$, $\left\langle E_{B}(\nu)\right\rangle$ and $\left\langle E_{tot}(\nu)\right\rangle$ are the time-averaged mechanical energy contents of $\textbf{u}\left(\textbf{r}\right)$ in region A, B and in the entire unit cell, respectively (see Fig.~\ref{SAW_solutions}(a)). The calculation is here performed starting from $\textbf{u}\left(\textbf{r}\right)$:
\begin{equation}
\alpha\left(\nu\right)=\frac{\int_{A}\rho\left(\textbf{r}\right)\textbf{u}^{2}(\textbf{r})d^{3}\textbf{r}+\int_{B}\rho\left(\textbf{r}\right)\textbf{u}^{2}(\textbf{r})d^{3}\textbf{r}}{\int_{tot}\rho\left(\textbf{r}\right)\textbf{u}^{2}(\textbf{r})d^{3}\textbf{r}}\;.
\label{SAW-likeness_equation}
\end{equation} 
The SAW-likeness coefficient $\alpha\left(\nu\right)$ outlines which eigenmodes of the system have mechanical energy mainly localized within a depth $\lambda$ (where $\lambda$ is both the nanostructure's period and the penetration depth of an unperturbed SAW of the same wavelength). In Fig.~\ref{SAW_likeness}, we report $\alpha\left(\nu\right)$ for three values of $p$. Two distributions, $\alpha_{g}\left(\nu\right)$ and $\alpha_{u}\left(\nu\right)$, emerge from the plot, corresponding to the cos-like and sin-like displacement profiles, respectively. We define pseudo-SAWs the eigenmodes with frequencies $\tilde{\nu}_g$ and $\tilde{\nu}_u$, corresponding to the two distributions' maxima. The pseudo-SAWs are the solutions that most likely resemble the unperturbed SAWs. The eigenfrequency degeneracy is removed and the opening of a frequency band gap is directly shown. We pin-point that the SAW-likeness coefficient is easily interpreted as the SAW line-shape, the line-shape concept arising because, in the composite system, the SAW is not an eigenmode.\\
\begin{table}[t]
\begin{center}
\centering \caption{\label{SAW_coeff} Calculated pseudo-SAW frequencies, corresponding SAW-likeness coefficients and linewidth $\gamma$ (FWHM) for both \textit{gerade} and \textit{ungerade} distributions.\\}
\begin{ruledtabular}
\begin{tabular}{llll}
\multicolumn{1}{c}{$p$} & \multicolumn{1}{c}{0.1} & \multicolumn{1}{c}{0.32} & \multicolumn{1}{c}{0.9} \\
\hline
$\tilde{\nu}_g$ (GHz) & 4.72 & 4.60 & 4.00 \\
$\alpha_g\left(\tilde{\nu}_g\right)$ & 0.38 & 0.12 & 0.70 \\
$\gamma_g$ (GHz) & 0.064 & 0.231 & 0.013 \\
\hline
$\tilde{\nu}_u$ (GHz) & 4.59 & 4.27 & 4.15 \\
$\alpha_u\left(\tilde{\nu}_u\right)$ & 0.18 & 0.07 & 0.73 \\
$\gamma_u$ (GHz) & 0.101 & 0.300 & 0.010 \\
\end{tabular}
\end{ruledtabular}
\end{center}
\end{table}
\indent In Fig.~\ref{SAW_likeness}(a) we report the calculation for a sample with $p=0.1$ ($d=100$~nm). For small filling fractions, nickel stripes act as a weak perturbation to the modes of the underlying substrate. The values of the pseudo-SAW frequencies and the corresponding SAW-likeness coefficients, as well as the linewidth $\gamma$ (FWHM) of the two distributions, are reported in Table~\ref{SAW_coeff}. The fact that $\alpha_{g}(\tilde{\nu}_g) > \alpha_{u}(\tilde{\nu}_u)$ means that the grating affects the sin-like SAW more than the cos-like one. Furthermore, the grating couples the unperturbed sin-like SAW over a wider range of modes: $\gamma_u~>~\gamma_g$. The frequency gap is $\Delta\tilde{\nu}=\tilde{\nu}_g-\tilde{\nu}_u=0.13$~GHz. Both $\tilde{\nu}_g$ and $\tilde{\nu}_u$ are lower than the unperturbed SAW eigenfrequency $\tilde{\nu}=4.92$~GHz.\\
\indent Increasing the filling fraction to $p=0.32$, we obtain the distributions reported in Fig.~\ref{SAW_likeness}(b). The physics is now well beyond the perturbative approach: larger linewidths and lower values of the SAW-likeness coefficient for the two pseudo-SAWs are the signatures of a stronger SAW-stripes interaction, as compared to the case of $p=0.1$. This leads to a higher coupling of surface waves with the bulk modes of the system, a confirmation to the evidence given in Fig.~\ref{SAW_solutions}(c) and Fig.~\ref{SAW_solutions}(d). The frequency gap is $\Delta\tilde{\nu}=0.33$~GHz. Both $\tilde{\nu}_g$ and $\tilde{\nu}_u$ are further down-shifted with respect to $\tilde{\nu}$. The pseudo-SAW frequency lowering is strongly dependent on the displacement field symmetry.\\
\indent Increasing the filling fraction to $p=0.9$, close to silicon full coverage (see Fig.~\ref{SAW_likeness}(c)), narrow linewidths and high peaked distributions resemble the frequency two-fold degenerate surface wave solutions shown for $p=1$ in Fig.~\ref{SAW_solutions}(e) and Fig.~\ref{SAW_solutions}(f), consistently with a reduced frequency gap $\Delta\tilde{\nu}=-0.15$~GHz. Interestingly $\Delta\tilde{\nu}$ changes sign and the \textit{ungerade} pseudo-SAW is now less affected by the periodicity as compared to its \textit{gerade} counterpart. For the case of $p=1$, $\tilde{\nu}$ is in agreement with the analytical theory developed by Auld.\cite{Auld}\\
\indent To further test the soundness of the pseudo-SAW definition adopted in the present work, we calculated for $p$= 0, 0.32 and 1, the dispersion for the first harmonic pseudo-SAW modes: $k_{x}\in[-0.2\pi/\lambda,\;0.2\pi/\lambda]$ and $n=1$. The outcome is shown in Fig.~\ref{SAW_dispersion}. The dispersion relation for the semi-infinite silicon slab ($p=0$) starts at $\nu=4.92$~GHz and is linear with a slope $v_{s}=4900$~m/s, in agreement with the speed of sound reported for SAW on Si(100).\cite{Auld} For the full overlay case ($p=1$), the dispersion starts at 4.12~GHz. Both solutions are true SAW eigenmodes of the system. For the case of $p=0.32$, we observe a gap opening and the two dispersion branches for the pseudo-SAWs remain correctly confined between their $p=0$ and $p=1$ analogues.\\
\section{Frequency Gap} \label{Frequency_gap}
\begin{figure}[t]
\centering
\includegraphics[bb= 85 417 515 773,keepaspectratio,clip,width=0.95\columnwidth]{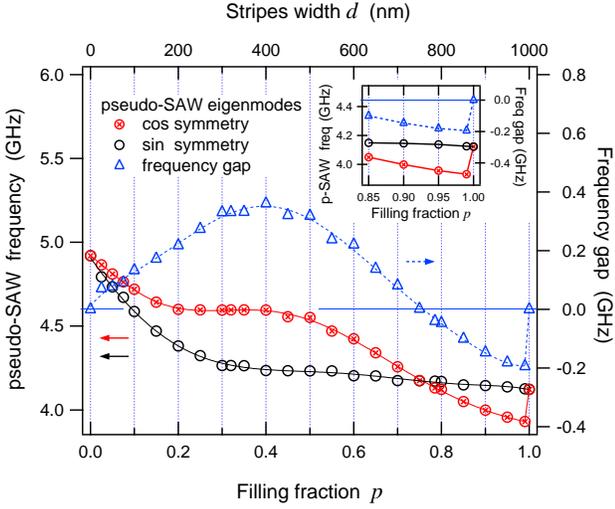}
\caption{(Color online) Left axis: pseudo-SAW eigenfrequencies $\tilde{\nu}_g$ (crossed circles) and $\tilde{\nu}_u$ (empty circles) versus surface filling fraction $p$ (or stripes width $d$ in nm, top axis). Right axis: pseudo-SAW frequency gap $\Delta\tilde{\nu}$ (triangles). The horizontal line is the $\Delta\tilde{\nu}=0$ line and refers to right axis only. Inset: magnification of the graph for high filling fractions $p$ is shown. Lines are a guide to the eye only.}
\label{Gap_filling}
\end{figure}

\indent Having formally defined the pseudo-SAW, we now address the issue of the frequency gap in the pseudo-SAW modes. In Fig.~\ref{Gap_filling} the pseudo-SAW frequencies and band gap calculated over the entire filling fraction range are shown. For $p\ne0$, the pure silicon slab eigenfrequency degeneracy is lifted. As $p$ increases, $\tilde{\nu}_g$ decreases until it reaches a plateau for $p=0.2$. The plateau lasts for $p\in\left(0.2,\;0.5\right)$ then, for $p>0.5$, $\tilde{\nu}_g$ decreases monotonically till it jumps back to the SAW value for full coverage as soon as the stripes' width $d$ reaches the size of the periodicity, becoming a complete overlay. There is a geometrical explanation for the discontinuity (see inset of Fig.~\ref{Gap_filling}). For $p=1$, periodic boundary conditions become effective also on the stripe's sides, the cos-like pseudo-SAW solution $\textbf{u}_{g}$ undergoes a transition from free to constrained displacement on the stripe's lateral boundaries. The same constraints do not affect $\textbf{u}_{u}$, because they are set where the nodes of the sin-like solution are found (see Fig.~\ref{SAW_solutions}(f)). In an experiment, for the above mentioned discontinuity, a steep positive derivative $\partial\tilde{\nu}_g$/$\partial$$p$ for $p\rightarrow1$ should be expected in place of an abrupt transition. Concerning $\tilde{\nu}_u$, it is a monotonically decreasing function of $p$. It decreases with a steep slope for $p<0.3$. It then changes slope stabilizing to an almost constant value all the way to $p=1$. For $p<0.75$, $\tilde{\nu}_g>\tilde{\nu}_u$; at $p=0.75$ the eigenvalues cross; for $p>0.75$, $\tilde{\nu}_g<\tilde{\nu}_u$; degeneracy is recovered for $p=1$. The frequency gap $\Delta\tilde{\nu}$ opens as soon as $p\ne0$. The maximum value $\Delta\tilde{\nu}=0.40$~GHz is attained when the silicon surface is close to half coverage. For higher filling fractions, the gap decreases to $\Delta\tilde{\nu}=0$ for $p=0.75$. For values of $p$ in excess of 0.75, $\Delta\tilde{\nu}<0$ and decreases till it makes an abrupt transition back to $\Delta\tilde{\nu}=0$ for $p=1$. The present results also show the limits of applicability of the perturbative approach.\cite{Robinson} Within a perturbative scheme the relative frequency shift is proportional to the filling factor: $(\tilde{\nu}(p)-\tilde{\nu}(0))/\tilde{\nu}(0)\propto(h/\lambda)p$. Calculation of the relative frequency shift, on the basis of the results reported in Fig.~\ref {Gap_filling}, shows that linearity holds up to $p\sim0.1$. For values of $p$ in excess of 0.1 one has to rely on the full calculations here reported.\\
\indent In the general case, for a fixed periodicity $\lambda$, the pseudo-SAW frequencies - both \textit{gerade} and \textit{ungerade}~- are functions of the parameters $\left\{p,h,m,\rho\right\}$: $\tilde{\nu}=f\left(p,h,m,\rho\right)$. The four parameters are not mutually independent, being $\rho\propto m/(hp)$. $\tilde{\nu}$ is then defined in a four-dimensional space and it is a function of three parameters only, the fourth being fixed by the choice of the other three. For instance, each data point reported in Fig.~\ref{Gap_filling} (left axis) is calculated having fixed $p$, $h$ and $\rho$ independently. The plots are then obtained spanning $p$ over the range $\left(0,1\right)$, thus exploring a particular trajectory $\tilde{\nu}(p,\;h=50$~nm,\;$\rho_{Ni})$ on the hyper-surface $\tilde{\nu}(p,h,\rho)$. A full description is computationally too time consuming and would not add much to the comprehension of the physics without slicing the hyper-surface along particular hyper-planes or trajectories.\\
\begin{figure}[t]
\centering
\includegraphics[bb= 125 91 523 563,keepaspectratio,clip,width=0.95\columnwidth]{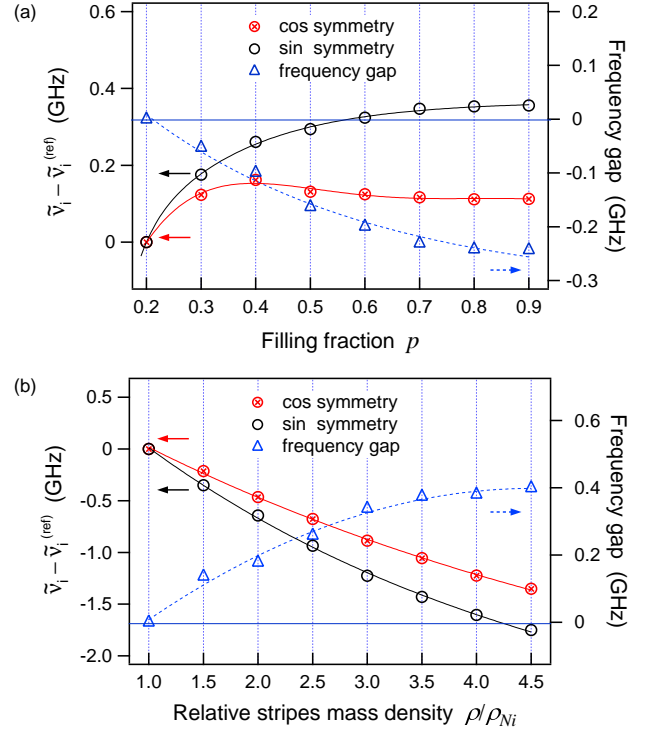}
\caption{(Color online) Geometric and mass loading effects on $\tilde{\nu}$. The reference system configuration is $p=0.2$, $h=50$ nm and $\rho=\rho_{Ni}$, hence $m$ is fixed to $m^{(ref)}$. The pseudo-SAW frequencies, calculated in the reference configuration, are $\tilde{\nu}^{(ref)}_g$ and $\tilde{\nu}^{(ref)}_u$. (a) Exploring the geometric effect: the mass loading factor $\left\{m,\rho\right\}$ is kept at the reference value, the variable being the geometric factor $\left\{p,h\right\}$. (b) Exploring the mass loading effect: the geometric factor $\left\{p,h\right\}$ is kept at the reference value, the variable being the mass loading factor $\left\{m,\rho\right\}$. Left axis: pseudo-SAW eigenfrequencies $\tilde{\nu}_g$ (crossed circles) and $\tilde{\nu}_u$ (empty circles). Right axis: pseudo-SAW frequency gap $\Delta\tilde{\nu}$ (triangles). The horizontal line is the $\Delta\tilde{\nu}=0$ line and refers to right axis only. Lines are a guide to the eye only.}
\label{Delta_nu}
\end{figure}
\indent To get a deeper physical inside, the effects on $\tilde{\nu}$ of the geometric factors $\left\{p,h\right\}$ and mass loading $\left\{m,\rho\right\}$, are here explored. We start considering a reference system configuration with $p=0.2$, $h=50$~nm and $\rho=\rho_{Ni}$, hence the value of $m$ is fixed to $m^{(ref)}$. The pseudo-SAW frequencies, calculated in the reference configuration, are $\tilde{\nu}^{(ref)}_g$ and $\tilde{\nu}^{(ref)}_u$. To test the contribution of the stripes' geometry, $m$ and $\rho$ are kept fixed at their reference values, while $p$ is allowed to increase from 0.2 to 0.9. Doing so, the value $h$ decreases. In Fig.~\ref{Delta_nu}(a) the dependences of $\tilde{\nu}_{i}-\tilde{\nu}^{(ref)}_{i}$ - with $i=\left\{g,u\right\}$ - and $\Delta\tilde{\nu}$ on the geometric factor $\left\{p,h\right\}$ are shown. The geometric factor up-shifts both frequencies, $\tilde{\nu}_u$ being more affected than $\tilde{\nu}_g$ and $\Delta\tilde{\nu}$ decreasing monotonically. The effect of the geometry in surface-based phononic crystal fits well with the results reported by Tanaka et al.\cite{Tanaka1998} for a 2D phononic crystal in which no mass loading effect is to be expected (the band gap definition is reversed in sign with respect to our case). This makes us confident that we are properly disentangling the geometry from mass loading effects. To inspect the mass loading effect, $p$ and $h$ are kept fixed at their reference values, while $\rho$ is raised up to 4.5~times the Ni density. Doing so, the value of $m$ increases. In Fig.~\ref{Delta_nu}(b) the dependences of $\tilde{\nu}_{i}-\tilde{\nu}^{(ref)}_{i}$ and $\Delta\tilde{\nu}$ on the mass loading factor $\left\{m,\rho\right\}$ are shown. The mass loading down-shifts both frequencies, $\tilde{\nu}_u$ being more affected than $\tilde{\nu}_g$ and $\Delta\tilde{\nu}$ increasing monotonically. The functions in Fig.~\ref{Delta_nu}(a) cannot be compared with the ones reported in Fig.~\ref{Delta_nu}(b), nor the absolute values are of relevance. The importance stands in the qualitative functions behavior, showing the disentangled physical effects of geometry and mass loading over $\tilde{\nu}_{i}$ and $\Delta\tilde{\nu}$. The present analysis shows that in a device the pseudo-SAW frequency gap results from a combination of geometry and mass loading factors, the two affecting the pseudo-SAW frequencies in opposite ways. A trade-off between $p$ and $m$ has to be foreseen for tailoring the pseudo-SAW frequencies or frequency gap in view of possible applications.
\section{Mechanical energy radiation} \label{Mechanical_energy}
\begin{figure}[t]
\centering
\includegraphics[bb= 84 77 496 406,keepaspectratio,clip,width=0.95\columnwidth]{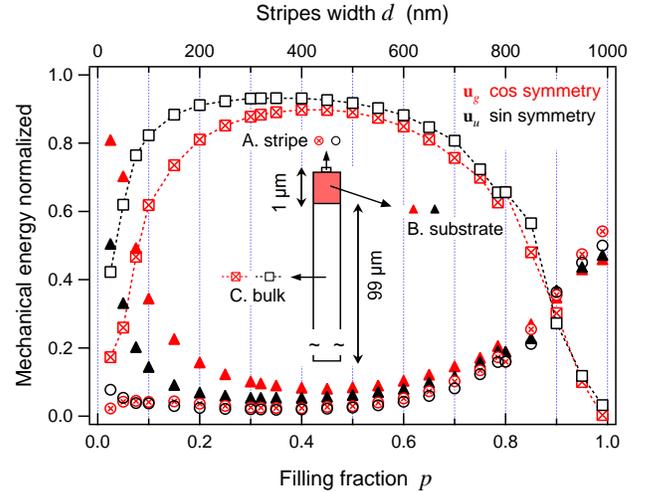}
\caption{(Color online) Analysis of the spatial distribution of the mechanical energy in the composite system over different filling fraction configurations, normalized against the total energy inside the cell. In the inset, the circles represent the fraction of the mechanical energy confined in the nickel stripes region A, the triangles are for the top 1~$\mu$m portion B of Si substrate and the squares are associated with the silicon bulk C. The dotted lines are a guide to the eye only.}
\label{Energy_filling}
\end{figure}

\indent In this section, the energy distribution in the system, for both pseudo-SAW eigenmodes $\textbf{u}_g$ and $\textbf{u}_u$, is studied as a function of $p$. Energy distribution in surface-based phononic crystal can be attained from the knowledge of the displacement or velocity and stress fields. The fields are typically calculated via coupled-mode theory or scattering theory in Born approximation. These are perturbative approaches. For instance, in this latter approach the SAW solution for the half-infinite Si slab is taken as the zero-order wave (impinging wave) used to evaluate the scattering matrix - the interaction term being brought in by the grating - together with final states modes (scattered waves). Such an approach loses reliability as the SAW frequency increases. Since the penetration depth of the SAW equals its wavelength $\lambda$, higher SAW frequencies imply stronger surface confinement and, ultimately, stronger scattering with the periodic grating. This requires considering higher order terms in the scattering integral equation in order to achieve a reasonably precise solution. For surface-based phononic crystals working in the hypersonic range ($\lambda\le1$~$\mu$m) the strong confinement is an issue. The theoretical frame here introduced to define the pseudo-SAW, together with the calculations of the composite system eigenmodes, can be exploited to tackle the problem of energy distribution in the system beyond a perturbative approach and the results can be readily translated in a scattering framework.\\
\indent The definition given in Eq.~\ref{SAW-likeness_coefficient} is here extended to account for the time-averaged normalized energy content in the three regions outlined in Fig.~\ref{SAW_solutions}(a) and in the inset of Fig.~\ref{Energy_filling}:
\begin{equation}
\alpha_{x}(\nu)\doteq\frac{\left\langle E_{x}(\nu)\right\rangle}{\left\langle E_{tot}(\nu)\right\rangle}\;,\ \ x=\left\{A,B,C\right\}\;.
\label{energy_distribution}
\end{equation}
\indent In Fig.~\ref{Energy_filling} the energy contents $\alpha_{x}(\tilde{\nu}_g)$ and $\alpha_{x}(\tilde{\nu}_u)$ are reported for $x=\left\{A,B,C\right\}$ as a function of the filling fraction $p$. The results show the mechanical energy spatial distribution - normalized against the total energy inside the cell - over the entire filling fraction range, when a pseudo-SAW eigenmode is excited. For $p=0$ the SAW is a true eigenmode, the energy being concentrated, as expected for a SAW, in region $B$: $\alpha_{B}(\tilde{\nu})\sim1$. For small filling fractions ($p<0.05$), at least for $\textbf{u}_g$, the elastic energy is mostly concentrated in the 1~$\mu$m top portion of the Si slab, where the unperturbed SAW is expected to dump. The grating is much more effective in scattering energy to the bulk for $\textbf{u}_u$: $\alpha_{B}(\tilde{\nu}_u)\sim\alpha_{C}(\tilde{\nu}_u)$. This symmetry-related difference in coupling a SAW to the bulk is a confirmation to the observed relationship $\gamma_u\sim2\gamma_g$ of Fig.~\ref{SAW_likeness}(a). As $p$ increases to half coverage, a greater amount of mechanical energy is transferred from the Si surface region B to the bulk region~C. From the eigenvalue problem perspective, a greater amount of energy radiated into the bulk translates in energy distributed over a wider range of eigenmodes of decreasing SAW-likeness coefficient $\alpha(\tilde{\nu})$, $\gamma$ increasing with $p$ for both \textit{gerade} and \textit{ungerade} pseudo-SAWs (see Fig.~\ref{SAW_likeness}(b)). The maximum values for both $\alpha_{C}(\tilde{\nu}_g)$ and $\alpha_{C}(\tilde{\nu}_u)$ are attained for $p=0.4$, the same filling fraction maximizing $\Delta\tilde{\nu}$. For $p>0.4$ the trend reverses, being $\partial\alpha_{C}(\tilde{\nu})/\partial p<0$ whereas $\partial\alpha_{A,B}(\tilde{\nu})/\partial p>0$; the relative energy content in the bulk region is transferred in region B and A. For $p\sim0.8$ the energy content in the Ni stripe equals the energy content in region B; the equality of energy content holds also between sin and cos symmetry eigenmodes and pseudo-SAW frequency crossing occurs (see Fig.~\ref{Gap_filling}). As the Si full coverage configuration is approached, the radiation into the bulk is further reduced in favor of a strong mechanical energy confinement in the stripe and in the top 1~$\mu$m portion of Si substrate, the crossing point being for $p=0.9$. For $p>0.9$, the oscillation of the Ni stripe starts to energetically dominate and the stripe behaves as a low-loss acoustic waveguide. The unperturbed SAWs, eigenmodes of the semi-infinite slab, are now heavily scattered into the Ni overlay. In the limit of Si full coverage ($p=1$), the Ni overlay acts as a lossless waveguide. The SAW is a true solution of the eigenvalue problem, without coupling with bulk modes, and it is frequency down-shifted with respect to the overlay-free SAW by an amount $\sim0.8$~GHz.
\section{Conclusions and Perspectives}
\indent In the present work we propose a theoretical frame allowing to access the physics of pseudo-surface acoustic waves in surface phononic crystals. The pursued strategy can be applied to any surface phononic crystal, enabling investigation of pseudo-SAW line-shapes, gap opening and mechanical energy scattering beyond perturbative approaches, thus finding application also in the hypersonic frequency range.\\
\indent We applied the outlined theoretical framework to the case of a hypersonic surface phononic crystals made of periodic nickel stripe on a silicon substrate. We then investigated the pseudo-surface acoustic wave frequency gap over the entire filling fraction range, starting from the case where the periodic nanostructures act as a perturbation, to finally reach the substrate full coverage. An understanding of how the construction parameters affect the frequency gap has been achieved. We showed that the pseudo-SAW frequency gap results from a combination of geometry $\left\{p, h\right\}$ and mass loading $\left\{m, \rho\right\}$ factors, the two affecting the pseudo-SAW frequencies in opposite ways. A trade-off between $p$ and $m$ is thus necessary for tailoring the pseudo-SAW frequencies or frequency gap for the application at hand. The mechanical energy spatial distribution of pseudo-SAW as a function of the filling fraction has been studied, allowing to tailor the device's parameters in order to minimize the energy content scattered out of the desired modes.\\
\indent The outlined theoretical scheme will proof a useful tool for future applications involving hypersonic pseudo-SAW in surface-phononic crystals. Complete control of the phonon gap opening in hypersonic surface-based phononic crystals, addressed in the present work, will be of impact in various fields of modern solid state physics. For instance the acoustic gap can be exploited to engineer surface waveguides and cavities to guide and localize hypersonic pseudo-SAWs. In addition, application of the recently demonstrated negative refraction effect\cite{Sukhovich} to surface phononic crystals, will foster the design of acoustic lenses working at hypersonic frequency. The possibility to focus and localize hypersonic pseudo-SAWs will enable efficient manipulation of the optical, electronic and magnetic properties of nanostructured devices in localized spatial regions. Finally, recent advances in the near-field optical microscopy pave the way to the coupling of the time-resolution of standard pump-probe optical experiments\cite{Giannetti, Comin} to the spatial resolution of SNOM probes.\cite{Courjon} This will allow to study in real-time the localization of hypersonic pseudo-SAWs in nanostructures or surface cavities with sub-wavelength spatial resolution.

\bibliography{Nardi_article}

\end{document}